\documentclass{PoS}

\title{Study of HERA data at low $Q^2$ and low $x$}

\ShortTitle{Higher twist at low $x$ in HERA data}

\author{\speaker{A M Cooper-Sarkar}\thanks{With thanks to I Abt, B Foster, M Wing, V Myronenko, K Wichmann}\\
        Oxford University, UK\\
        E-mail: \email{amanda.sarkar@cern.ch}}

\abstract{In the HERAPDF2.0 PDF analysis it was noted that the fit $\chi^2$ worsens significantly at low $Q^2$ for both NLO and NNLO fits. The turn over of the reduced cross section at low-$x$ and low $Q^2$ due to the contribution of the longitudinal cross section $F_L$ is also not very well described. In this paper the prediction for $F_L$ is highlighted and the corresponding extraction of $F_2$ from the data is further investigated, showing discrepancies with description of HERAPDF2.0 at low $x$ and $Q^2$.  The effect of adding a simple higher twist term of the form ~$F_L*A/Q^2$ to the description of $F_L$ is investigated. This results in a significantly better description of the reduced cross-sections, $F_2$ and $F_L$ at low $x$, $Q^2$ and a significantly lower $\chi^2$ for the NNLO fit as compared to the NLO fit. This is not the case if the higher twist term is added to $F_2$}

\FullConference{XXIV International Workshop on Deep-Inelastic Scattering and Related Subjects\\
		11-15 April, 2016\\
		DESY Hamburg, Germany}

\begin{document}

HERA data may shed light on low-$x$ physics and the transition to the 
non-perturbative regime at low $Q^2$. Ever since the rise of $F_2$ at low-$x$ was observed it has been 
speculated that there may be a need for QCD resummations beyond the conventional DGLAP equations, whether
 these be $ln(1/x)$ resummations a la BFKL or the need for non-linear evolution equations which take account 
of gluon recombination and the possibilty of gluon saturation.
The final combined data on NC and CC $e^+p$ and $e^-p$ inclusive cross sections from H1 and ZEUS are now published ~\cite{newcomb}. These data were used as the input for
next-to-leading order (NLO) and next-to next to 
leading order (NNLO) QCD PDF fits resulting in a PDF set called
 HERAPDF2.0. The HERAPDF2.0 analysis used the RTOPT heavy flavour scheme~\cite{robert} as default and the fits 
described in the present contribution are done in this scheme unless otherwise stated. 
It has been observed that the $\chi^2$ of the QCD fits in the DGLAP formalism is worse at low $Q^2$. 
The kinematic reach of HERA is such that low $Q^2$ is also low $x$. 
It has been suggested that in this kinematic region diagrams with two, three and four gluons in the 
t-channel could give rise to higher twist terms which contribute to the longitudinal structure function $F_L$ while cancelling between the longitudinal and transverse components in $F_2$~\cite{bartels}.
The NC $e^+p$ data includes data at 
different centre-of-mass energies such that different values 
of $y$ are accessed at the same $x,Q^2$. This gives information on the longitudinal structure function $F_L$. The
 present contribution looks at the QCD fits concentrating on the role of the longitudinal structure function $F_L$.
Further details are given in Ref.~\cite{hht}

\begin{figure}[tbp]
\begin{center}
\begin{tabular}{cc}
\includegraphics[width=0.5\textwidth]{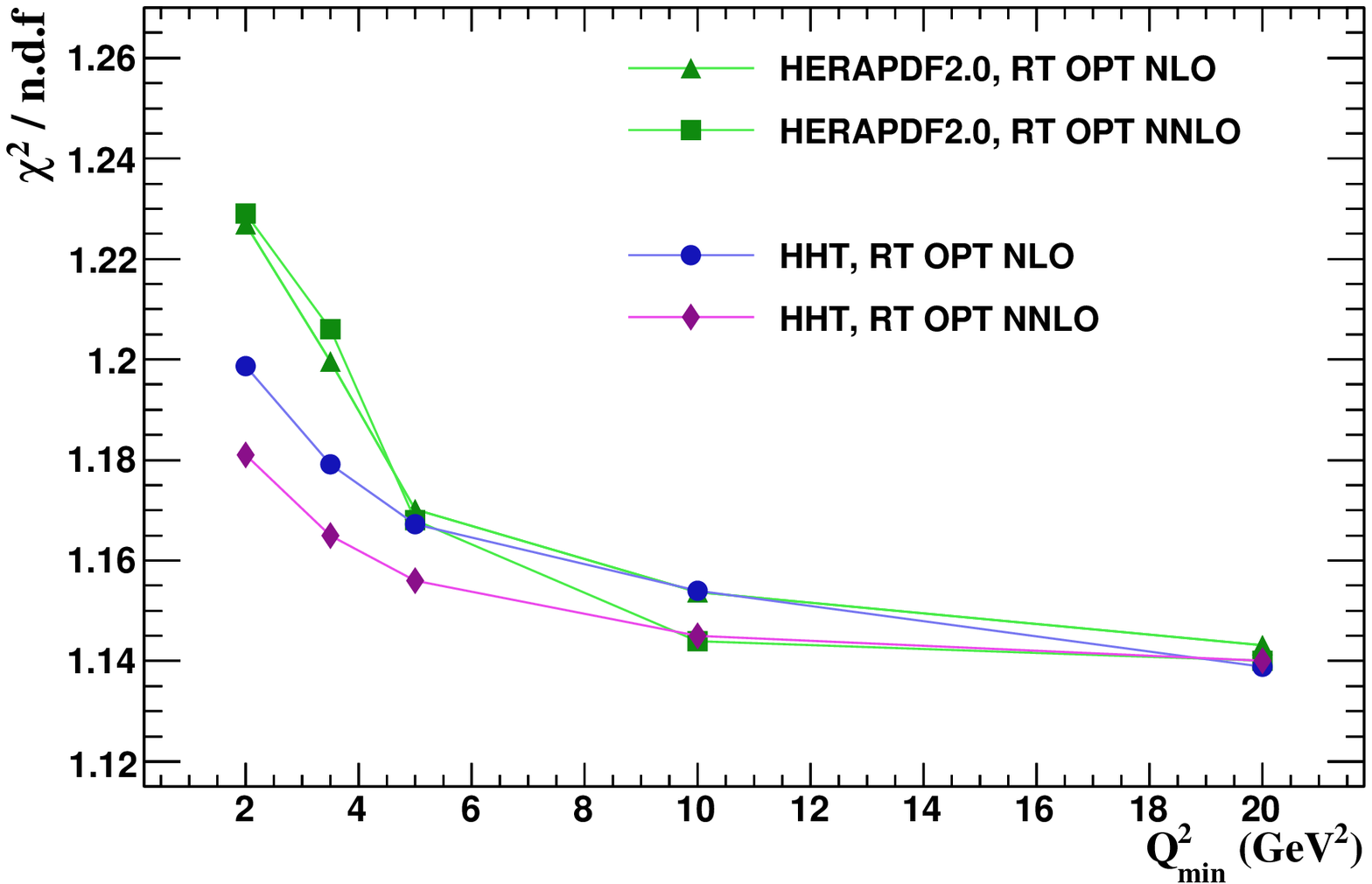} &
\includegraphics[width=0.5\textwidth]{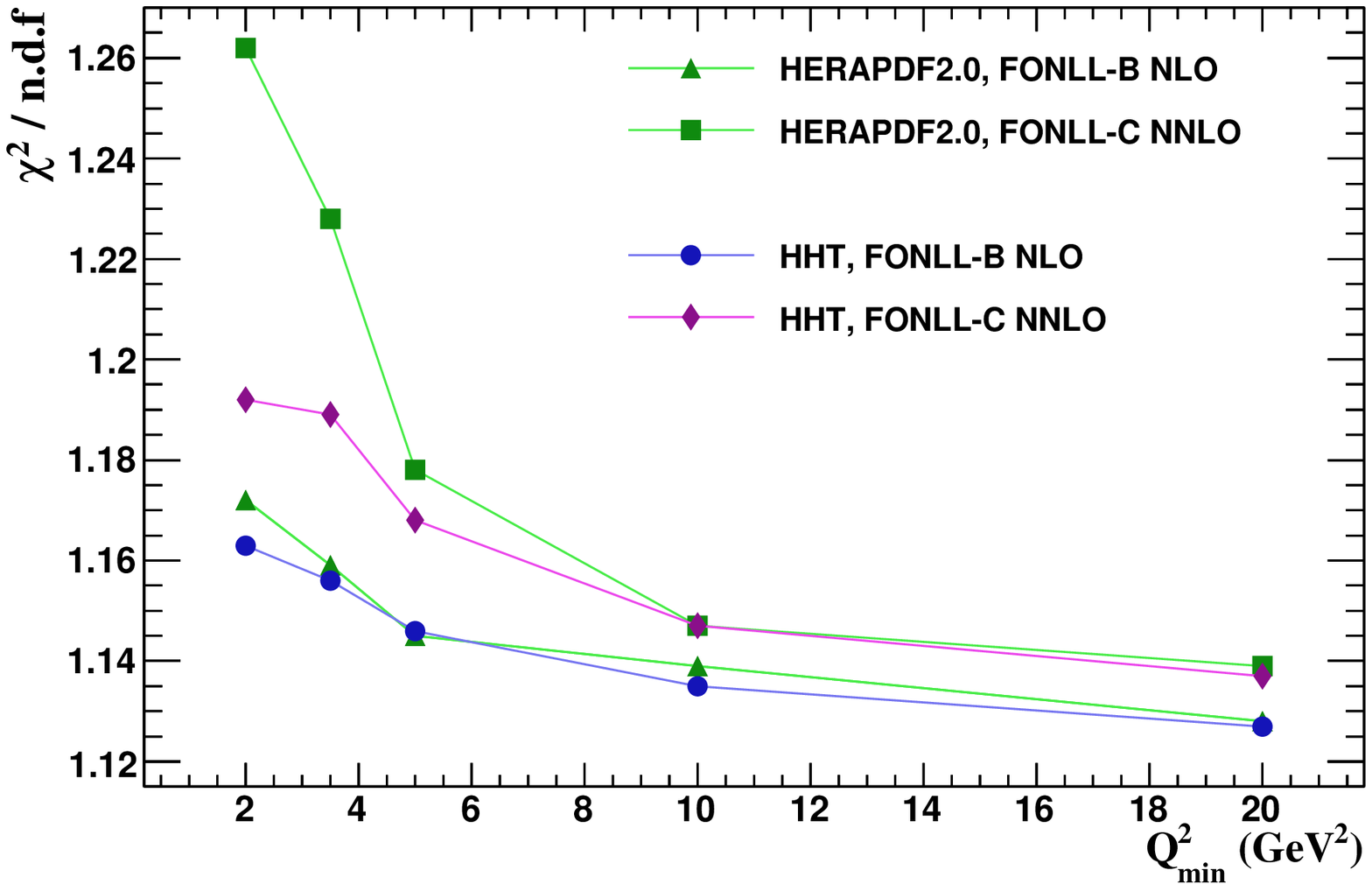}
\end{tabular}
\end{center}
\caption { 
The $\chi^2$ per degree of freedom vs the minimum $Q^2$ of data entering the 
HERAPDF2.0 NLO and NNLO fits and the corresponding HHT fits which include a higher twist term in $F_L$.
Left: using the default RTOPT heavy flavour scheme. Right: using the FONLL heavy flavour scheme.
}
\label{fig:chiscan}
\end{figure}

The dependence of the $\chi^2$ on the $Q^2$ cut applied to the data was investigated in the 
HERAPDF2.0 analysis, and the results are shown in  
Fig.~\ref{fig:chiscan} which shows
$\chi^2$ per degree of freedom vs the minimum $Q^2_{min}$ 
of data used in the fit, for the HERAPDF2.0 NLO and NNLO QCD fits. 
It is interesting to see whether the situation can be improved by modification of the leading twist QCD 
predictions for $F_L$ by a simple higher twist term such that 
$F_L(HT) = F_L*(1 + A/Q^2)$.
The fits of the HERAPDF2.0 analysis are repeated using this modified $F_L$ with $A$ as a free parameter. 
All the other conditions are set as for HERAPDF2.0. These fits are called the HHT QCD fits.
One can see a very significant decrease in $\chi^2$ particularly at NNLO, such that
the NNLO fit is now significantly better than the NLO fit. 
The values of $A$ extracted are quite high: $A=4.2\pm0.7$~GeV$^2$ for NLO and $A=5.5\pm0.6$~GeV$^2$ for
 NNLO fits. Details of the $\chi^2$ are given in Table.~\ref{tab:chitab}
\begin{table}
\begin{center}
  \begin{tabular}{llll}
 \hline
 \hline
   Type of fit $Q^2_{min} = 3.5$GeV$^2$ &  HERAPDF2.0  &  HHT & $A_{HT}$ \\
\hline
  NNLO $\chi^2$/ndof        &  $1363/1131$          &  $1316/1130$ &$5.5\pm0.6$\\ 
$\chi^2$/ndp for NC$e^+p$:$Q^2> Q^2_{min}$      &  $451/377$               & $422/377$&\\
$\chi^2$/ndp for NC$e^+p$: $2.0$~GeV$^2 < Q^2 < Q^2_{min}$ & $41/25$ & $32/25$&\\
\hline
 NLO $\chi^2$/ndof        &  $1356/1145$          &  $1329/1145$ &$4.2\pm 0.7$ \\
$\chi^2$/ndp for NC$e^+p$:$Q^2> Q^2_{min}$     &  $447/377$               & $431/377$&\\
$\chi^2$/ndp for NC$e^+p$: $2.0$~GeV$^2 < Q^2 < Q^2_{min}$ & $46/25$ & $46/25$&\\
 \hline
 \hline
\end{tabular}
\end{center}
\caption{Table of $\chi^2$ per degree of freedom (ndof) for HERAPDF2.0 and HHT fits both with $Q^2_{min}=3.5$~GeV$^2$. 
Also given are the $\chi^2$ per number of data points (ndp) for the high precision NC $e^+p$ data at $\sqrt{s}=318$~GeV for $Q^2> Q^2_{min}$. The final row in each 
category represents the $\chi^2$ per number of data points for predictions of the fits below the fitted 
region, from $Q^2 = 3.5$ to $Q^2 = 2.0$. In addition the values of the higher twist parameter $A$ are given for the HHT fits.}
\label{tab:chitab}
\end{table}

In the default RTOPT heavy flavour scheme $F_L$ is calculated to 
O($\alpha_s^2$) at NLO and at O($\alpha_s^3$) at NNLO. The behaviour of the $\chi^2$ as $Q^2_{min}$ is raised can 
differ somewhat according to the heavy flavour scheme used ~\cite{newcomb}. In particular the $\chi^2$ for 
low $Q^2$ data is lower for schemes in which $F_L$ is calculated to O($\alpha_s$). The HERAPDF2.0 and HHT fits 
have also been performed in the FONLL~\cite{fonll} schemes B and C in which $F_L$ is calculated respectively 
to O($\alpha_s$) at NLO and at 
O($\alpha_s^2$) at NNLO. The results are also shown in Fig.~\ref{fig:chiscan}.
The NNLO fit behaves in much the same way as the fits in the RTOPT scheme, with a much decreased $\chi^2$ for the 
HHT fit. The NLO fit does not need a large higher twist term essentially because an $F_L$ calculation at  O($\alpha_s$) already produces a larger $F_L$. However as soon as $F_L$ is calculated to O($\alpha_s^2$) or higher the need for a higher twist term appears.

\begin{figure}[tbp]
\vspace{-0.3cm} 
\begin{center}
\includegraphics[height=0.3\textheight]{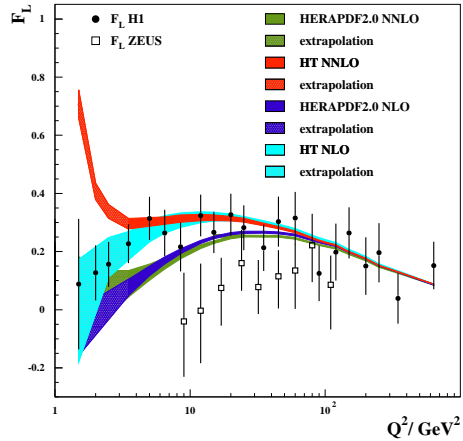}
\end{center}
\caption {The separate H1 and ZEUS measurements of $F_L$ compared to the HERAPDF2.0 NLO and NNLO QCD fit predictions and to the corresponding predictions of the HHT fits which include a modified higher twist term in $F_L$.
}
\label{fig:flht}
\end{figure}

\begin{figure}[tbp]
\begin{center}
\begin{tabular}{cc}
\includegraphics[width=0.5\textwidth]{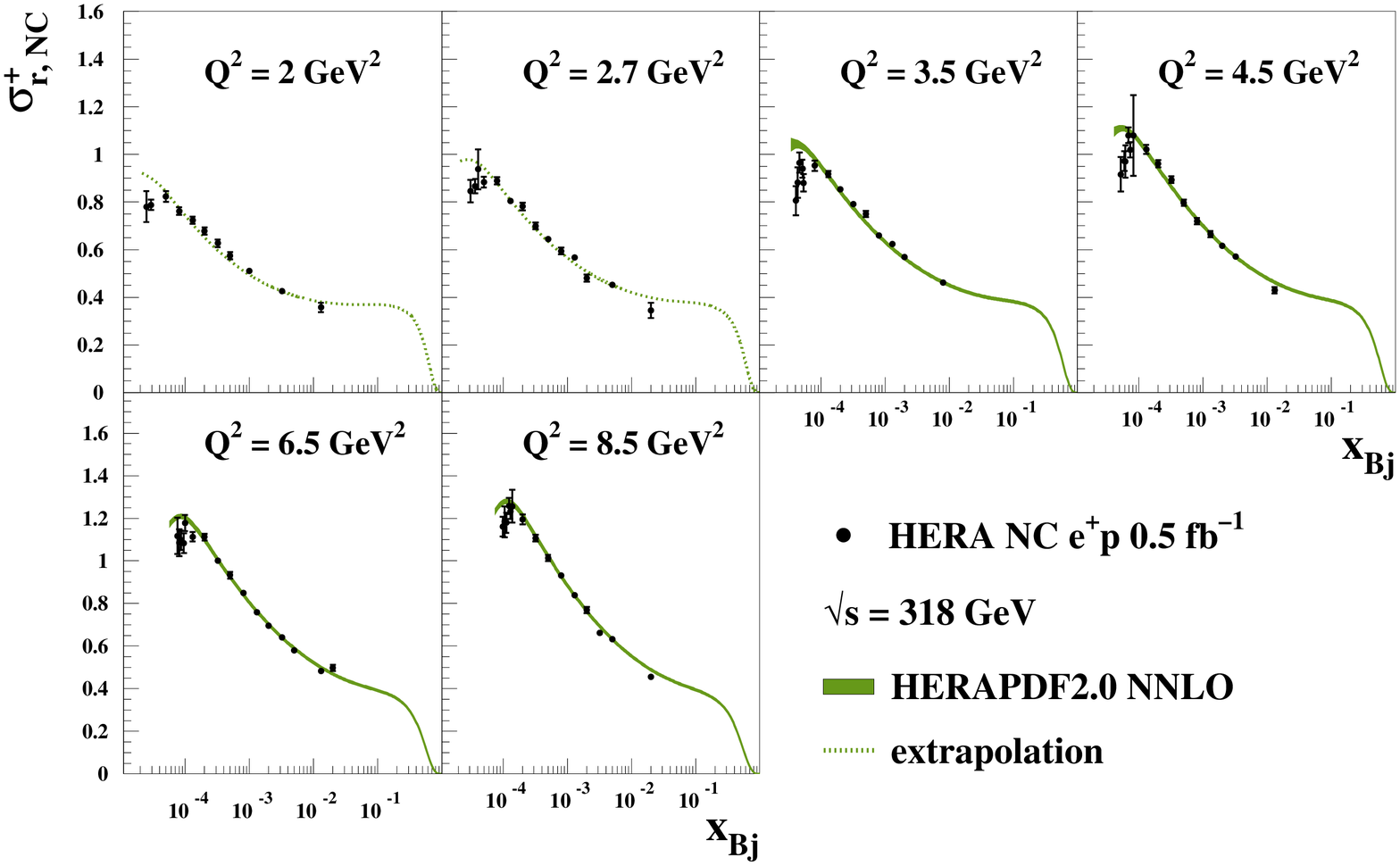} &
\includegraphics[width=0.5\textwidth]{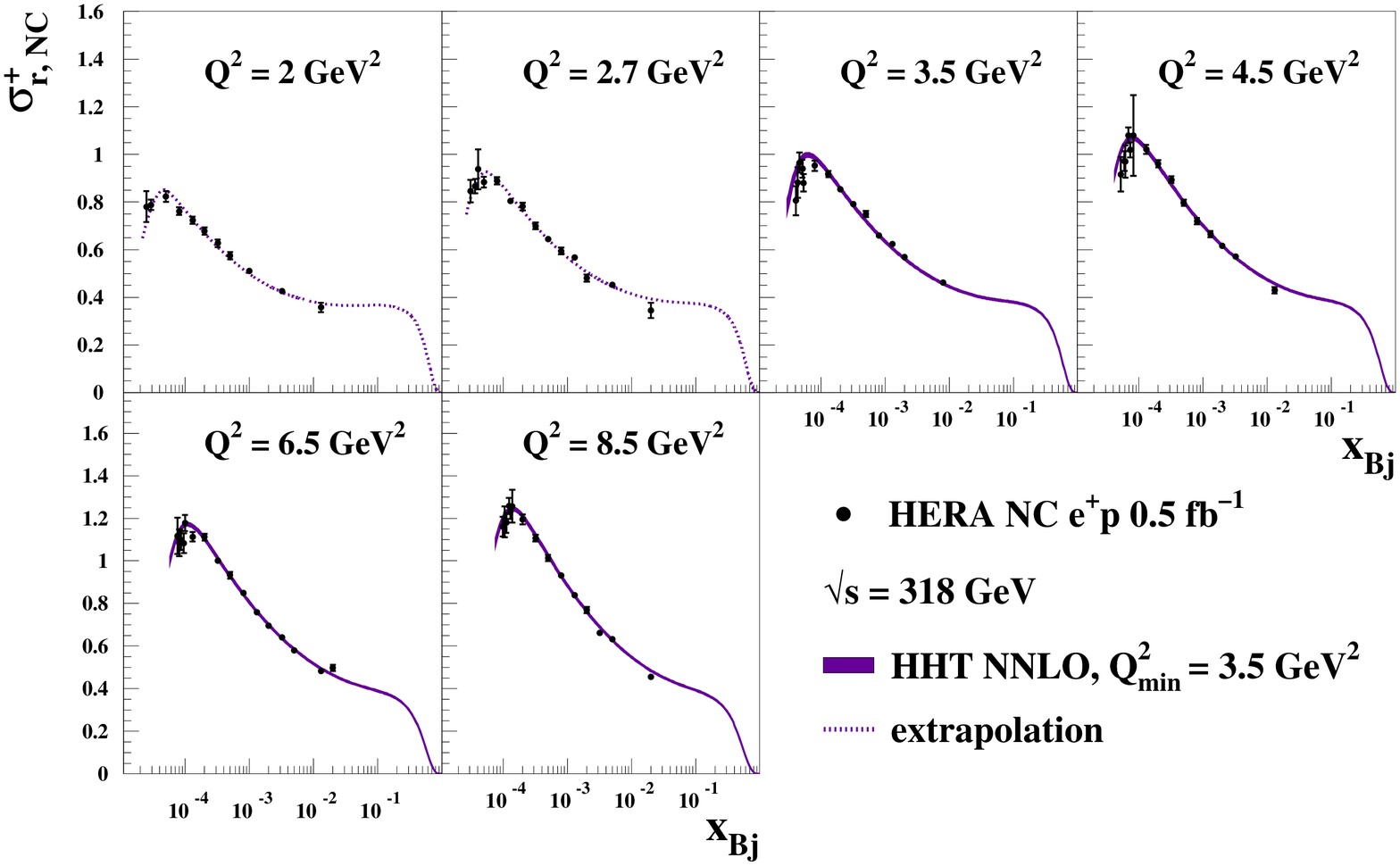}
\end{tabular}
\end{center}
\caption {The combined HERA measurements of $\sigma_{reduced}$ compared to the predictions of HERAPDF2.0 NNLO(left)
 and the corresponding predictions of HHT NNLO(right). 
}
\label{fig:sight}
\end{figure}

Fig.~\ref{fig:flht} shows the HERAPDF2.0 and HHT predictions for $F_L$  
superimposed 
on the separate H1 and ZEUS measurements. 
The corresponding 
predictions for the reduced cross-section, $\sigma_{red} = F_2 - y^2/Y_+ F_L$, $Y_+=1+(1-y^2)$, for HERAPDF2.0 and HHT at NNLO are shown in Fig.~\ref{fig:sight} (NLO results are similar).
A clear improvement in the description of the data is 
seen for the fits including higher twist. This is mostly due to the improved description of the turn-over at low $x$ which comes from an increased $F_L$. 

\begin{figure}[tbp]
\begin{center}
\begin{tabular}{cc}
\includegraphics[width=0.5\textwidth]{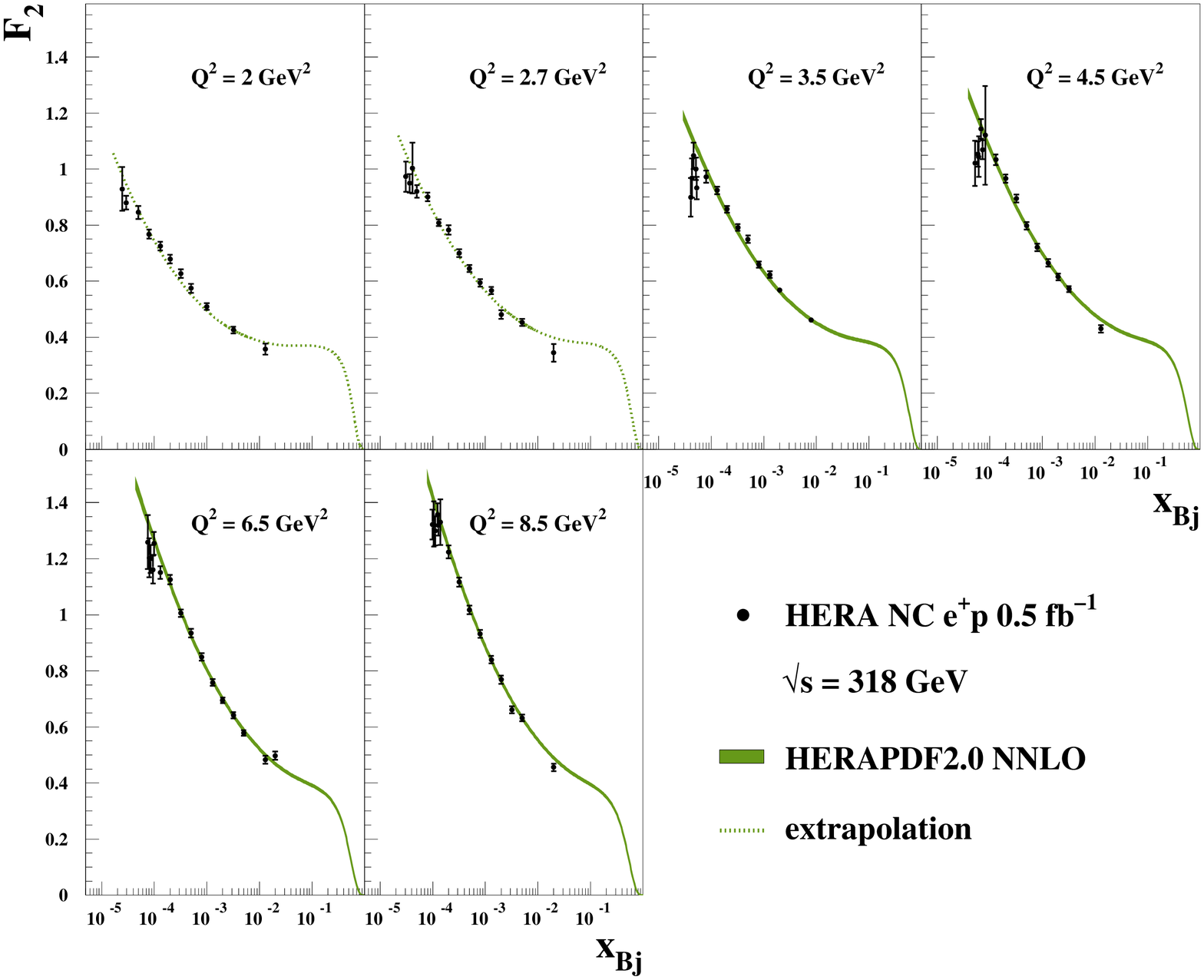} &
\includegraphics[width=0.5\textwidth]{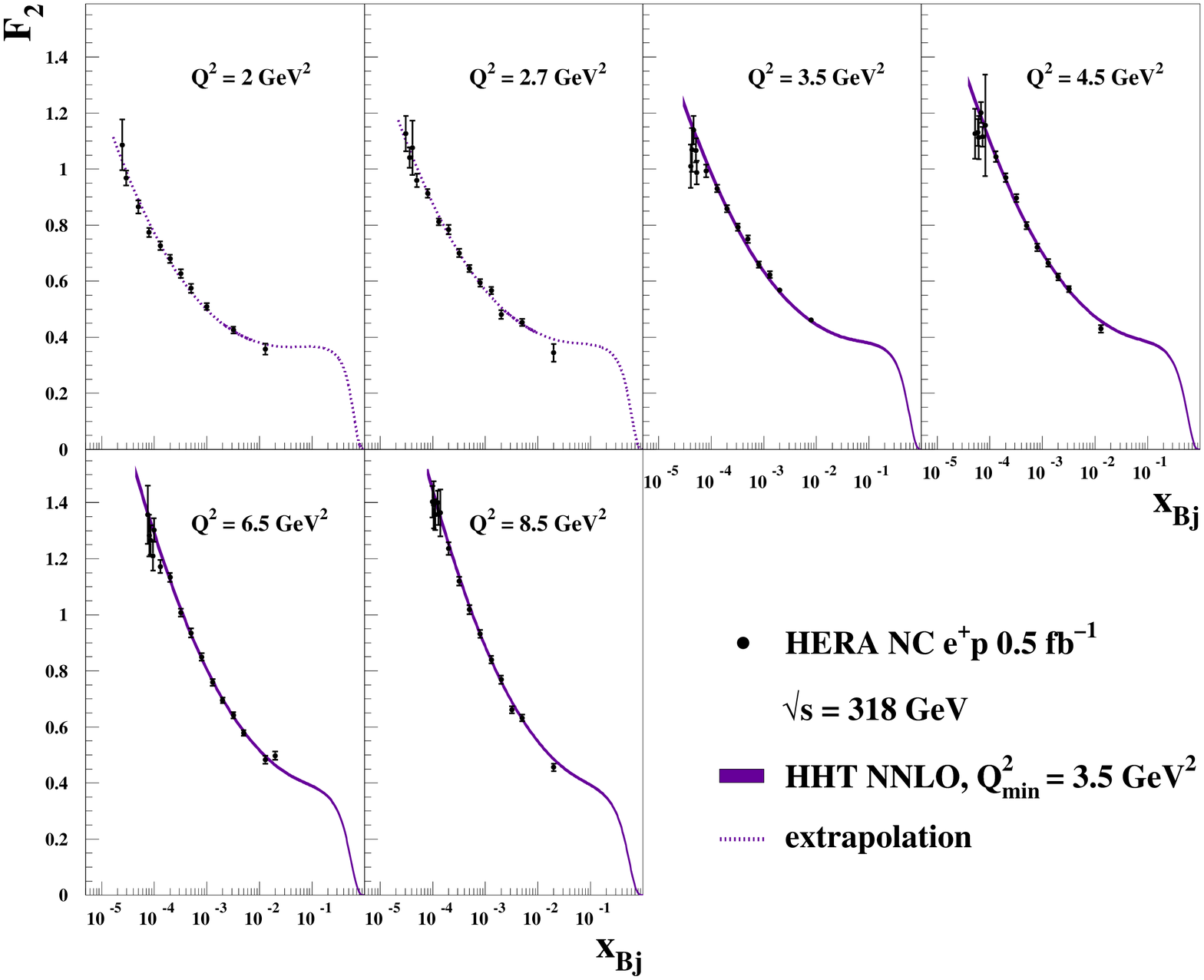}
\end{tabular}
\end{center}
\caption {The combined HERA measurements of $F_2$ compared to the predictions of HERAPDF2.0 NNLO(left) and the 
corresponding predictions of  HHT NNLO(right).
}
\label{fig:f2ht}
\end{figure}
\begin{figure}[tbp]
\begin{center}
\begin{tabular}{cc}
\includegraphics[width=0.5\textwidth]{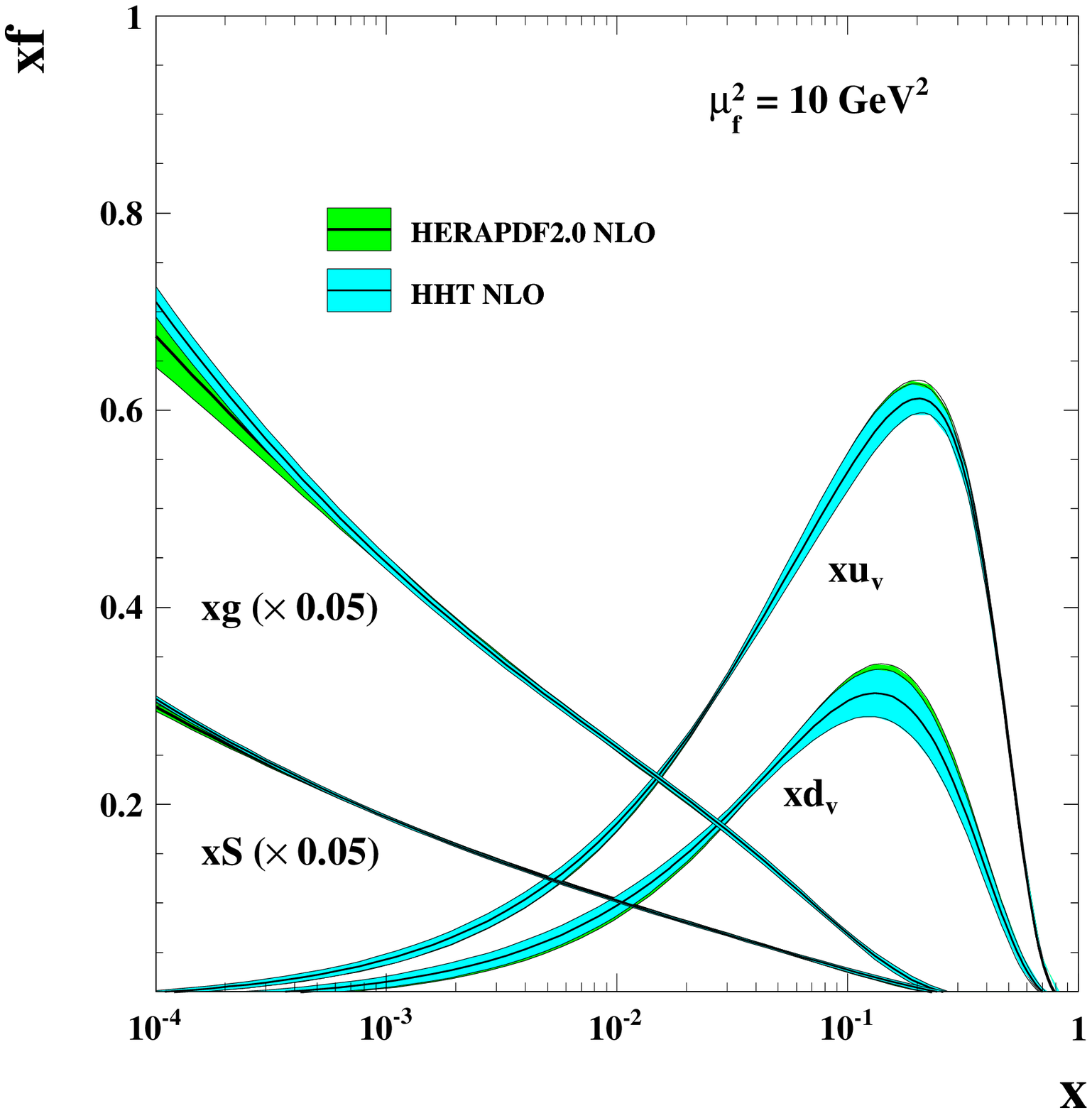} &
\includegraphics[width=0.5\textwidth]{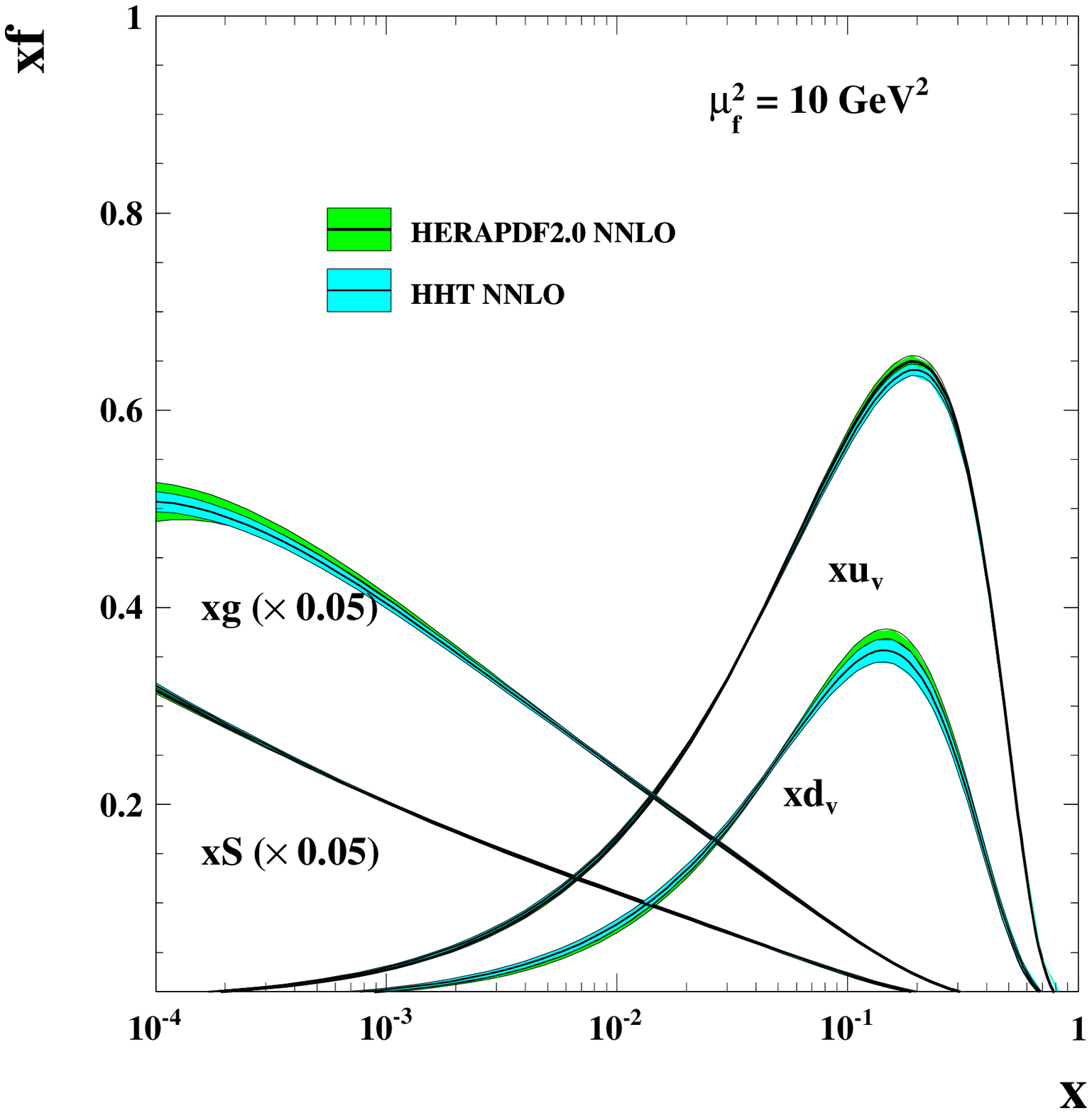}
\end{tabular}
\end{center}
\caption {The PDFs resulting from the HHT fits compared to those of HERAPDF2.0 at NLO(left) and at NNLO(right).
The gluon and sea distributions are scaled down by a factor of 20. Experimental uncertainties on both fits are shown}
\label{fig:pdfs}
\end{figure}

It is also interesting to look at the predictions and extracted data for $F_2$, see Fig.~\ref{fig:f2ht}. 
Measurements of $F_2$ are extracted as $F_2^{extracted} = F_2^{predicted} *\sigma_{red}^{measured}/\sigma_{red}^{predicted}$.
Since $F_2$ is a dominant part of the cross section this is a reasonable procedure, but clearly if $\sigma_{red}^{measured}$ is lower than $\sigma_{red}^{predicted}$ then $F_2^{extracted}$ will also be low and the consequence for the
HERAPDF2.0 $F_2$ extraction is that $F_2^{extracted}$ itself starts to take a turn-over at low-$x$, low$Q^2$ which is not 
in agreement with QCD predictions for $F_2$. However, for the HHT fit this problem is mitigated, the $F_2^{extracted}$ 
does not take a significant turn-over and is in much better agreement with the predictions. Note that the 
predictions for $F_2$ are very similar for the two fits since these depend only on the PDFs and not on the 
higher twist term. The PDFs extracted from the HHT fits are very similar to the HERAPDF2.0 PDFS as shown in Fig.~\ref{fig:pdfs}.
In particular, despite the fact that a good fit requires a larger $F_L$ contribution at low-$x$ and low-$Q^2$,
 the NNLO fit still requires a gluon parametrisation with a substantial negative term such 
that it the gluon starts to turn over at low-$x$ and $Q^2$. Using an alternative parametrisation without this term 
(such that the gluon is always positive definite above the starting scale of evolution) 
results in much higher $\chi^2$ for both the HERAPDF2.0 fit and the HHT fit.

We have also investigated the inclusion of a higher twist term in $F_2$ such that $F_2(HT) = F_2*(1 + A/Q^2)$.
In this case there is only a small improvement in $\chi^2$ and the value of 
$A$ is also small, consistent with zero: $A =0.12\pm 0.07$GeV$^2$. If higher twist terms are inlcuded in both 
$F_2$ and $F_L$ the result is similar to including the term only for $F_L$. 

The fits shown in Figs.~\ref{fig:sight},\ref{fig:f2ht} were done for $Q^2> 3.5$~GeV$^2$ as for HERAPDF2.0.
However it can be seen that the predictions of the HHT fit describe the data well down to $Q^2=2.0$~GeV$^2$. 
Thus new HHT fits were performed with $Q^2_{min}$=2.0$~GeV^2$. 
The details of the $\chi^2$ are given in Table.~\ref{tab:chitab2}
\begin{table}
\begin{center}
  \begin{tabular}{llll}
 \hline
 \hline
   Type of fit $Q^2_{min} = 2.0$GeV$^2$ &  HERAPDF2.0  &  HHT & $A_{HT}$ \\
\hline
  NNLO $\chi^2$/ndof        &  $1437/1171$          &  $1381/1170$ &$5.2\pm0.7$\\ 
$\chi^2$/ndp for NC$e^+p$:$Q^2> Q^2_{min}$      &  $486/402$               & $457/402$&\\
$\chi^2$/ndp for NC$e^+p$: $2.0$GeV$^2 < Q^2 < Q^2_{min}$ & $31/25$ & $26/25$&\\
\hline
 NLO $\chi^2$/ndof        &  $1433/1171$          &  $1398/1170$ &$4.0\pm 0.6$ \\
$\chi^2$/ndp for NC$e^+p$:$Q^2> Q^2_{min}$     &  $487/402$               & $466/402$&\\
$\chi^2$/ndp for NC$e^+p$: $Q^2> Q^2_{min} < Q^2 < 2.0$GeV$^2$ & $40/25$ & $31/25$&\\
 \hline
 \hline
\end{tabular}
\end{center}
\caption{Table of $\chi^2$ per degree of freedom (ndof) for HERAPDF2.0 and HHT fits both with $Q^2_{min}=2.0$~GeV$^2$. 
Also given are the $\chi^2$ per number of data points (ndp) for the high precision NC $e^+p$ data at $\sqrt{s}=318$~GeV for $Q^2> Q^2_{min}$. The final row in each 
category represents the $\chi^2$ per number of data points for predictions of the fits from $Q^2 = 3.5$ to $Q^2 = 2.0$. In addition the values of the higher twist parameter $A$ are given for the HHT fits.}
\label{tab:chitab2}
\end{table}

\begin{figure}[tbp]
\begin{center}
\includegraphics[height=0.3\textheight]{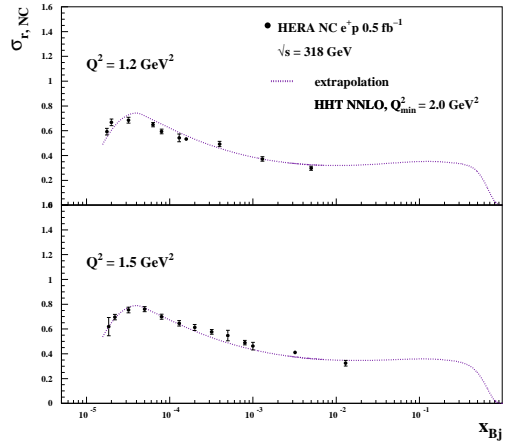}
\end{center}
\caption { 
The predictions of the HHT fit done for $Q^2_{min}= 2.0$GeV$^2$, for lower $Q^2$ data.
}
\label{fig:sight1.2}
\end{figure}

The fit quality for the data points in the range 
$2.0 < Q^2 < 3.5$~GeV$^2$ improves somewhat particularly at NLO. However the fitted parameters are much the same as for the fit with $Q^2_{min}=3.5$~GeV$^2$. In particular the extracted values of the higher twist 
parameter $A$ are almost the same. Looking at the predictions for $\sigma_{red}$ for even lower 
$Q^2$ values, see Fig.~\ref{fig:sight1.2}, shows that the description is apparently 
good even down to $Q^2 = 1.2$~GeV$^2$. However if we look at the corresponding predictions for $F_L$, 
see Fig.~\ref{fig:flht}, it is evident that this simple description is not tenable for $Q^2 < \sim 2.0$~GeV$^2$.

In summary, the introduction of a higher twist term in the description of the longitudinal structure function $F_L$, 
in the context of pQCD fits within the DGLAP formalism,  significantly improves the description of HERA data on deep inelastic scattering at low $x$ and low $Q^2$ down to 
$Q^2 < \sim 2.0$~GeV$^2$. The introduction of a similar term in the structure function $F_2$ is unnecessary 
confirming the expectation that higher twist terms cancel between the longitudinally and transversely polarised 
photons in $F_2$. However further mechanisms are necessary to describe data for $Q^2 < \sim 2.0$~GeV$^2$.


\begin{thebibliography}{99}
\bibitem{newcomb} H.~Abramowicz et al, Eur. Phys. J75(2015)580 , arXiV:1506.06042
\bibitem{bartels} J.~Bartels, K.~Golec-Biernat and H.~Kowalski, Phys Rev D66(2002)14001, 
\bibitem{hht} I.~Abt et al, arXiV:1603.02299.
\bibitem{robert} R~.S~.Thorne and R.~G.~Roberts, Phys.ReV.D57(1998)6871; R.S.Thorne, Phys.Rev.D73(2006)054019; R.S. Thorne, Phys.Rev.D86(2012)074017
\bibitem{fonll} S.~Forte et al, Nucl.Phys.B834(2010)116

\end{thebibliography}
\end{document}